# A diamond heater-thermometer microsensor for measuring localized thermal conductivity: a case study in gelatin hydrogel

*Linjie Ma[#], Jiahua Zhang[#], Zheng Hao, Jixiang Jing, Tongtong Zhang, Yuan Lin, Zhiqin Chu\**

Linjie Ma, Jiahua Zhang, Jixiang Jing, Tongtong Zhang

Department of Electrical and Electronic Engineering, The University of Hong Kong, Hong Kong

Zheng Hao, Yuan Lin

Department of Mechanical Engineering, The University of Hong Kong, Hong Kong

Zhiqin Chu

Department of Electrical and Electronic Engineering (Joint Appointment with School of Biomedical Sciences), The University of Hong Kong, Hong Kong

E-mail: zqchu@eee.hku.hk

[#] These authors made equal contributions to this work.

Understanding the microscopic thermal effects of the hydrogel is important for its application in diverse fields, including thermal-related studies in tissue engineering and thermal management for flexible electronic devices. In recent decades, localized thermal properties, such as thermal conductivity, have often been overlooked due to technical limitations. To tackle this, we propose a new hybrid diamond microsensor that is capable of simultaneous temperature control and readout in a decoupled manner. Specifically, the sensor consists of a silicon pillar (heater) at about 10 microns in length, topped by a micron-sized diamond particle that contains silicon-vacancy (SiV) centers (thermometer) with $1.29 \text{ K} \cdot \text{Hz}^{-\frac{1}{2}}$ temperature measurement sensitivity. Combining this innovative, scalable sensor with a newly established simulation model that can transform heating-laser-induced temperature change into thermal conductivity, we introduced an all-optical decoupled method with about $0.05 \text{ W} \cdot \text{m}^{-1}\text{K}^{-1}$ precision, which can reduce laser crosstalk. For the first time, we track the thermal conductivity change of hydrogels during the gelation process and demonstrate the existence of variation. We introduce a rapid, undisturbed technique for measuring microscale thermal conductivity, potentially serving as a valuable tool for cellular thermometry and highlight the idea that decoupling can reduce crosstalk from different lasers, which is helpful for quantum sensing.

## 1. Introduction

Microscale thermal conductivity between cells and their microenvironment is crucial for cellular metabolism, [1-2] tissue development, [3-4] and disease treatment. [5] Specifically, in the field of tumor hyperthermia, [6] detecting the thermal conductivity between cancer cells and their microenvironment at the single-cell level has significant challenges. It on the one hand requires to improve the spatial resolution and accuracy of micro and nanoscale thermal measurement techniques. On the other hand, a deeper understanding of the thermal properties of ECM protein gels used to mimic the cellular microenvironment in vitro is highly demanded. [7] However, the localized thermal conductivity of ECM-based biomaterials has not been studied well due to the lack of efficient measurement methodologies. [8]

The ideal hydrogel is a homogeneous system; in most experiments, the macroscopic properties of the hydrogel are often used to characterize it. However, the study of the microscopic mechanical properties of hydrogels shows the existence of spatial variations in crosslinking density, which induce differences in microstructure and stiffness, [9]further influencing cellular behavior. [10] The crosslinking density of the hydrogel also influences its macroscopic thermal conductivity, [11] but their microscopic influence is not clear. By far, most of the existing thermal conductivity studies involving hydrogels focus on large-scale thermal conductivity [12-13] instead of micro-scale. Several methods have been developed to improve the spatial resolution. One of them is scanning probe-based thermometry. [14-16] However, it can only get the property of the surface layer of the sample. Another type is heater-thermometer nanosensor. It can also enable thermometry into sub-diffraction limited spatial resolution, [17] and has been used to perform thermal conductivity measurements [18], but the overlapping of the heating laser and sensing laser may induce crosstalk. With limited experimental support, there are only a few theoretical and simulation studies focused on the micro-/nano scale. [19] What is currently desired is a microscale thermal conductivity measurement method that can be easily integrated into existing systems without introducing crosstalk for measurement, while combining high sample shape adaptability and low time consumption.

Color centers hosted in the diamond lattice are new choices for thermal-related measurement. Diamond is famous for its extremely high thermal conductivity, great hardness, excellent photostability, [20] and remarkable chemical stability. [21-22] Several studies have illustrated spin-based thermometry [23-24] and all-optical thermometry [25-26] rely on the color center existing in the diamond. Compared with spin-based thermometry, although there is a compromise in the sensitivity, [22] free of microwave excitation is a great advantage of the all-optical method, which not only reduces the complexity of the measurement hardware but also eliminates the heating effect of the microwave [27]. In particular, one of the color centers, the silicon-vacancy center (SiV), has become a promising candidate for thermometry as it offers good properties such as narrow zero-phonon line (ZPL) bandwidth and more concentrated power on the fluorescent spectrum. [26] Compared with other all-optical thermometry based on NV centers, such as ZPL-based [28-29] and lifetime-based [30] methods, it will provide better sensitivity. [22, 31] The all-optical detection capabilities and optical-based rapid readout abilities endowed by color centers in diamonds provide a foundation for overcoming the problems in the current measurement of the thermal conductivity of hydrogels. Here, we present a novel sensor utilizing diamond-based all-optical thermometry. The sensor consists of an approximately ten-micrometer-long silicon pillar, with a micrometer-sized diamond particle containing ensemble SiV Centers positioned at one of its tops, which provides high sensitivity ($\sim 1.29 \text{ K} \cdot \text{Hz}^{-\frac{1}{2}}$). Based on the new sensor, we proposed an all-optical decoupled measurement strategy (**Figure 1**). In contrast to conventional measurement methods, the readout laser and heating laser are spatially separated and directed towards opposite ends of the sensor, thereby minimizing crosstalk between the heating and sensing lasers. Also, it has been demonstrated that the heating laser can induce a maximum local temperature up to 70 kelvins, [23] such high temperature will cause the denature of most of the ECM proteins. By splitting the two lasers, we can minimize the potential negative impacts caused by the high-power heating laser. A simulation model is built based on the morphology of the sensor to derive the localized thermal conductivity from the measured rising of temperature induced by the heating laser. Due to the high sensitivity in temperature measurement, the accuracy of thermal conductivity measurement can reach ~0.05 $\text{W} \cdot \text{m}^{-1}\text{K}^{-1}$ after calibration. With the help of the new method, we track the microscale thermal conductivity of hydrogels during the gelation process and demonstrate the existence of variation.

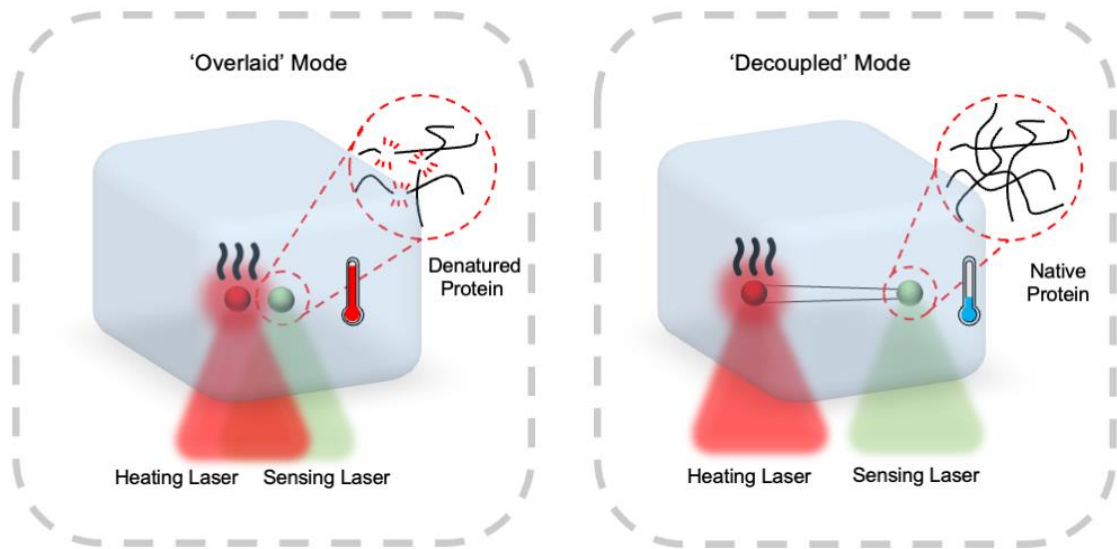

**Figure 1.** Schematic of traditional sensing methods of thermal conductivity (heater-thermometer hybrid nanostructure) and decoupled sensing method. The heating laser is a 637nm high power laser with milliwatt power. The sensing laser is a 532nm laser with microwatt power.

## 2. Result and Discussion
## 2.1. Fabrication of the Sensor

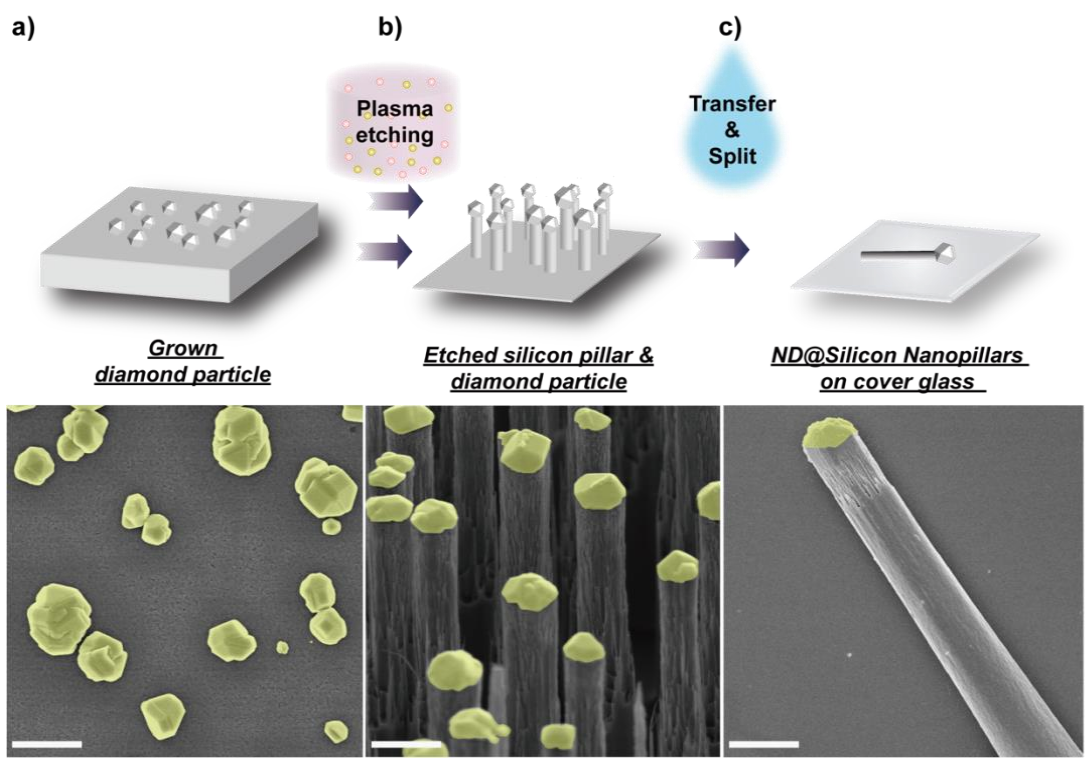

**Figure 2.** Fabrication and characterization of ND@silicon nanopillars. **a.** TEM of grown diamond particle. **b.** SEM of Silicon pillar after inductively coupled plasma (ICP) etching. **c.** SEM of pillar after transferred to the surface of the cover glass. The scale bars are 2μm. Diamond particles are painted with yellow.

A specialized and easily fabricated sensor is essential for the implementation of our newly proposed thermal conductivity measurement approach. We designed the sensor exclusively in a pillar shape with a single diamond particle at the top, hosting silicon-vacancy (SiV) centers to achieve all-optical thermometry and thermal conductivity measurements. Using our recently developed fabrication scheme, [32] we successfully synthesized nano/micro diamonds with ensemble SiV centers. Firstly, salt-assisted air-oxidized (SAAO) [33] nanodiamonds were deposited onto a silicon wafer using spin-coating, serving as seeds for subsequent chemical vapor deposition (CVD) to grow larger diamond particles **(Figure 2a)**. We then applied the plasma etching to sculpt the silicon substrate, with the grown diamond particles acting as a mask to yield silicon pillars **(Figure 2b)**. These pillars were later transferred onto a cover glass and separated into individual. Each sensor is crowned with a diamond particle on top **(Figure 2c)**. The sensor's morphology was characterized using an atomic force microscope (AFM). The silicon pillar and the diamond particle can be clearly seen and distinguished in **Figure 3a**. The AFM line profile reveals that the sensor size is approximately 13.1 μm in length and 2.6 μm in width (refer to Supplementary Information). We can customize the geometry and density of the pillars to suit various application needs by modulating experimental conditions during the spin coating, CVD processing, and plasma etching.

More specially, the parameters of spin coating affect the final density and distribution of the sensor on the silicon wafer. Longer CVD processing time makes diamond particles grow larger. As they serve as the mask, it's obvious that a larger particle size will make a larger sensor in diameter. The length of the sensor is mainly determined by the time of plasma etching. Longer etching time causes more silicon to be removed from the uncovered substrate, resulting in a greater length of the pillar. Sensor fabrication with multi-parameters is detailed in the supplementary information (SI) section.

## 2.2. Temperature Sensitivity

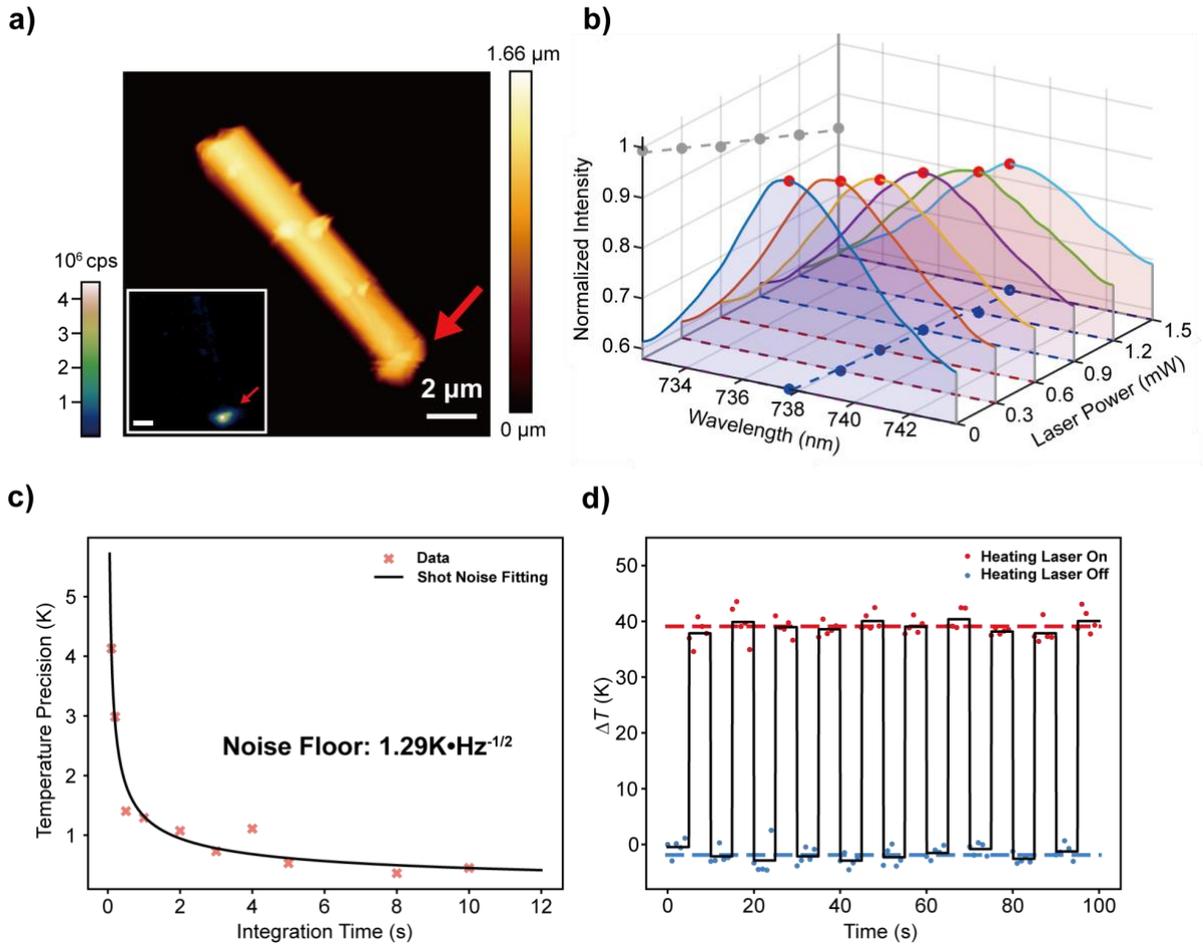

**Figure 3.** Decoupled heating and temperature measurement. **a.** AFM measurement of the pillar and the corresponding confocal fluorescence image (insert). The red arrow indicates the location of the diamond particle. The scale bars in the figure are 2μm. **b.** Spectrum under different laser power in the air. **c.** Shot noise measurement under 500μW illumination laser power. **d.** Heating laser on and off loop for 10 loops, each loop is 10 seconds, each point is a measurement result of 1-second integration time.

To verify the existence and characteristics of SiV centers in the sensor, we utilized confocal microscopy to measure its optical properties. The confocal scanning image, depicted as an insert in Figure 3a, elucidates the distribution and intensity of fluorescence. The red arrow distinctly marks the location of the diamond particle, where the fluorescence intensity is observed to increase to approximately 4 million counts per second. Spectral analysis of this illuminated region unveiled a signature SiV center spectrum, thereby corroborating the presence of SiV centers, as detailed in SI. The rest part of the sensor, the silicon pillar, shows almost no fluorescence signal.

The accuracy of temperature measurement plays a crucial role in the sensor designed for thermal conductivity measurement, as it significantly impacts the overall precision of the result.

We perform the shot noise measurement for the single particle using a confocal microscope and get a sensitivity of 1.29 $K \cdot Hz^{-\frac{1}{2}}$ (**Figure 3c**), giving a temperature precision under 1K at 2 second integration time. Compared to detonation nanodiamonds (DND)-grown diamonds which have a 3.42 $K \cdot Hz^{-\frac{1}{2}}$ sensitivity, [32] our sample shows a better performance.

Our experimental results indicate that increasing the integration time does not significantly enhance measurement precision when it is larger than 1 second. The measurement precision is not only limited by the inherent sensitivity of the sensor, but also the equipment noise caused by the spectrometer. Based on the shot noise theory, extending the integration time will increase the precision. However, increased dark noise accumulation, which is the main contribution to the noise at long integration times, will limit the precision. Using a spectrometer with deep-cooling capabilities could improve outcomes for prolonged integration periods. However, maintaining consistent mechanical stability over extended periods presents a challenge. Moreover, time-consuming experiments might not be suitable for dynamic scenarios as they require rapid temporal resolution. [34]

A more strategic approach would be to augment the photon capture rate within a specified duration. Increasing the number of SiV centers in the diamond particles to enhance the photoluminescence can effectively improve sensor efficiency. The method we used to produce the diamond particles on our sensor has been proven to have much better properties, such as significantly increased photoluminescence and better spectrum quality than conventional DND-grown diamonds. [32] It provides our sensor with higher sensitivity.

## 2.3. All-Optical Decoupled Sensing Method

With our latest design and fabrication of the sensor, we introduce a distinctive all-optical decoupled thermal conductivity sensing technique using a confocal microscope. This method employs two lasers: a 532 nm green laser to excite the SiV centers and a 637 nm red laser to heat the silicon pillar, [35] both directed into the objective. The green laser is focused on a fixed point, specifically at the center of the objective's field of view, consistent with traditional confocal microscopy. A piezo stage is used for sample scanning and aligning the diamond particle with the laser spot. The focal point of the red laser is adaptable. By adjusting the incident angle of the laser, the focal point can be maneuvered within a constrained area. This allows the red laser to be focused on the opposite end of the silicon pillar, thereby raising the sensor's temperature. The rising amplitude of the sensor is related to the thermal conductivity of the surrounding material. By monitoring the rise in temperature, the localized thermal conductivity can be deduced.

In most conventional all-optical thermometry and photothermal heating techniques, the illumination and heating functions often focus at a single focal point, [36] or a single laser plays both roles, [18] as most sensors and heaters are designed as nano/micro-sized spheres. However, these techniques have the inherent disadvantage that the heating location and the sensing location cannot be separated. When the focal points of two lasers coincide, crosstalk can occur: the heating laser may disturb the fluorescence characteristics of the thermometer, and the measurement laser might inadvertently cause photoheating. Additionally, if a single laser is used for both heating and illumination, changes in fluorescence intensity may occur when modulating the heating power.

Our innovative decoupled sensing methodology avoids these issues by using a pillar-shaped sensor and separating the focal points of the two lasers on different segments of the sensor. This configuration permits the independent adjustment of each laser without mutual interference. This novel decoupled sensing paradigm holds the potential to introduce new methodologies across various sensing domains. For example, in spin-based quantum sensing, conventional techniques may not work, because introducing an additional laser could influence the spin state. The decoupled sensing approach can eliminate this challenge, as the controlled spin remains outside the illumination range of the second laser.

Besides that, the sensor's unique shape, combined with the new decoupled sensing method, provides several advantages. First, the distance between the sensor and the thermometer is under control. By changing the condition of fabrication, the sensors with different lengths can be manufactured (detailed in SI). It allows us to tune the size of the sensor based on the requirements of different applications. Second, the distance between the sensor and the thermometer remains fixed under all conditions. In future experiments, if the sensor is used for living cells, we do not need to worry about the heater's and sensor's relative movement. Our method provides us with a robust way to put a heater at a designed distance away from the sensor.

In our experiment, we directed a green laser at the diamond particle to excite the SiV centers, while simultaneously aiming a red laser at the opposite end of the silicon pillar to increase the sensor's temperature. The intensity of the heating laser ranged from 0 to 1.5 mW. Theoretically, as the power of the red laser increases, the sensor's temperature should rise due to the photothermal effect of the silicon. [35] This theoretical prediction is confirmed by spectral measurements, which reveal changes in both the ZPL position and bandwidth, indicating a temperature rise as the heating laser's power increases. (see **Figure 3b**). We further assessed the stability of this laser heating technique (see **Figure 3d**). The heating laser cycled through 5 seconds-on and 5 seconds-off cycles for a total of 100 seconds. Throughout the "on" phase, the laser's power remained constant. Remarkably, the sensor's temperature demonstrated consistent stability during both the laser's on and off phases. This evidence suggests that our approach is both reversible and highly stable.

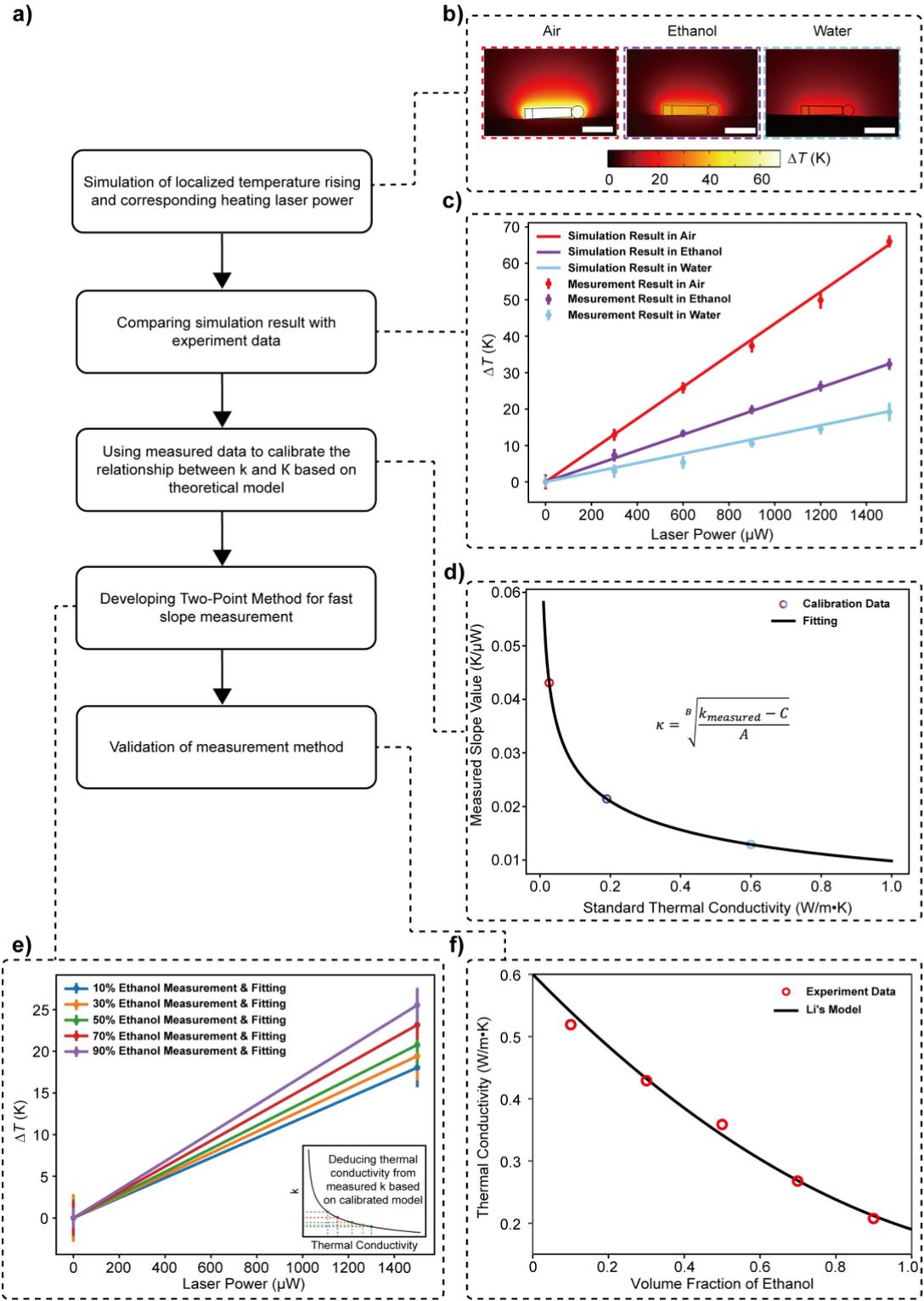

**Figure 4.** Thermal conductivity extraction using a combined experimental and simulation approach. **a.** Flow chart of extracting thermal conductivity from temperature measurement. **b.** Simulation figures of temperature distribution in water, ethanol, and air. The scale bar is 5μm. **c.** The simulation result and the measurement result of temperature change with laser power in water, ethanol, and air. **d.** The relationship between k and thermal conductivity. **e.** The measurement result of temperature change with laser power in ethanol solution with different concentrations. Insert shows deducing thermal conductivity of ethanol solution from a calibrated model. **f.** Comparing the measured and theoretical thermal conductivity of ethanol solution.

## 2.4. Reveal Thermal Conductivity from Temperature-changing

A numerical simulation model has been developed to reveal the relationship between the temperature rise induced by the heating laser and the thermal conductivity of the adjacent material (detailed in SI). We developed a geometric model of the sensor based on its dimensions derived from the photo. The segment of the silicon pillar illuminated by the red laser generates heat due to the photothermal effect, so we designated it as the heat source. Heating power is calculated based on laser power. For different parts of the sensor, we assign different thermal conductivity based on the intrinsic properties of the material so the thermal dispersion can be accurately simulated. Simulations were conducted in three different media—air, ethanol, and water—arranged in order of increasing thermal conductivity. The temperature distribution resulting from the simulation is shown in **Figure 4b**. The findings suggest that the rising amplitude of the temperature decreases as the thermal conductivity of the surrounding material increases.

Measurements were conducted in each of the three materials separately. In each material, we vary the intensity of the heating laser and record the ZPL shift in the SiV spectrum. Subsequently, the ZPL shift is translated into an increase in temperature. According to our theoretical calculations, within the same material, the temperature increase and laser intensity exhibit a linear relationship (see SI for details). **Figure 4c** displays both the theoretical simulation results and the data collected from the experiment. We can see from the figure that the experiment results align well with the simulation. Additionally, the slope of the $\Delta T$-Laser Power curve, denoted as $k$, varies among different materials.

Furthermore, through theoretical analysis, we found that the slope has a relationship with the thermal conductivity of the surrounding material, denoted as $\kappa$. This relationship can be expressed as $k = A * \kappa^B + C$. The data obtained from measurements in three different materials were employed to ascertain the unknown parameters in this equation, serving as the calibration procedure for the sensor. Subsequently, we established the correlation between the slope $k$ and the thermal conductivity $\kappa$ (as illustrated in **Figure 4d**). With this calibrated model, we can deduce the thermal conductivity of the material from the measured slope, $k$, of the $\Delta T$ versus the heating laser power curve.

To validate our method, we use the calibrated sensor to measure ethanol solutions of varying concentrations. After the calibration, the slope is the only thing we need to know for deducing thermal conductivity. In the previous calibration measurement, we have demonstrated that the measured data shows a linear relationship between temperature rise and heating laser power. As the slope can be determined with only two points, we can optimize the measurement method to reduce the time required for measurement by performing a 'Two-Point Measurement Method', which only measures two points to get the slope. The measured data demonstrated in **Figure 4e** shows differences in the value of $k$. By substituting the measured values of $k$ into the model (**Figure 4e insert**), we obtain the corresponding thermal conductivity for each solution. The theoretical thermal conductivity of a mixture of two liquids can be calculated using Li's model. [37] We compare the experimental data with the theoretical calculations (**Figure 4f**). Both sets of results demonstrate high consistency, and the error analysis demonstrates an accuracy of ~$0.05 W \cdot m^{-1} K^{-1}$ (see SI).

## 2.5. Dynamic Thermal Conductivity Tracking During Gelation

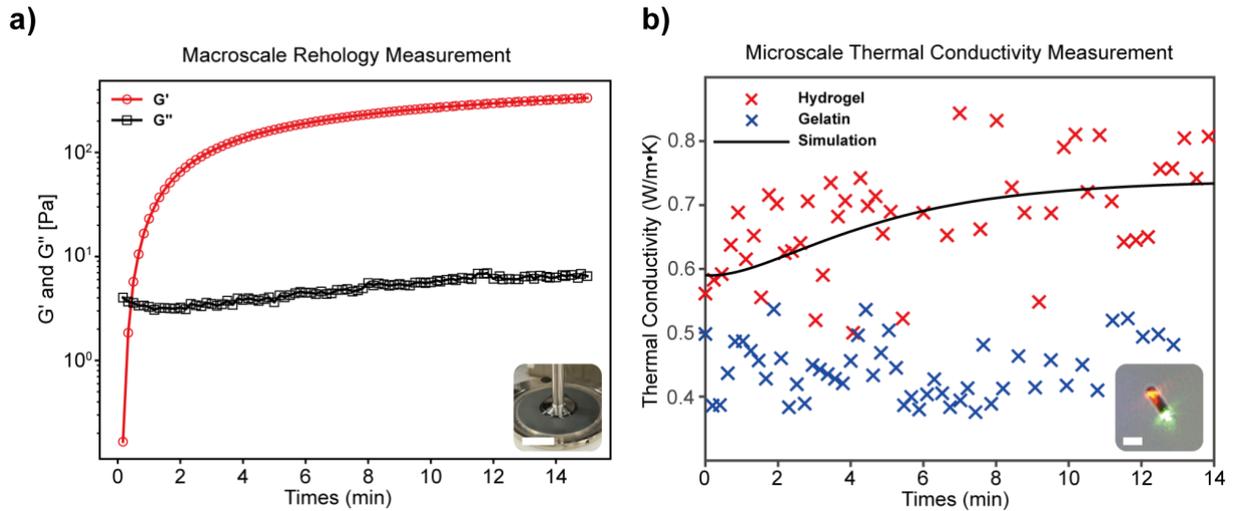

**Figure 5.** In-situ thermal conductivity Simulation and measurement. a. Macroscale rheology of the Gelatin hydrogel gelation process. G' is storage moduli and G'' is loss moduli. Insert shows the hydrogel sample for the rheology test. The scale bar is 25mm. b. The measurement result of microscale thermal conductivity in gelatin solution and during gelatin hydrogel cross-linking

process. Simulation shows the change of thermal conductivity in the hydrogel cross-linking process. Insert is the sensor under the microscope view. The scale bar is 10μm.

Hydrogel is a widely used material in cell and tissue engineering. In the past decades, it has been proven that cells can be mechanically stimulated, [38] and the mechanical properties of the extracellular matrix can influence cell behavior. [39] Nowadays, cell-level heat transfer and stimulation are gaining more and more attention. Most research focuses on intracellular [18] and intercellular [40] thermal phenomena. However, little attention has been paid to the possible influence of the extracellular matrix. One reason is the lack of proper measurement methods. In addition, hydrogel, which is a widely used extracellular matrix, is a highly dynamic material, making its measurement challenging.

With the newly developed method for measuring thermal conductivity, the issue has been addressed. We applied it to localized thermal conductivity measurements on a hydrogel, consisting of a gelatin solution and glutaraldehyde. Significant changes were observed in both the loss modulus and storage modulus during the gelation process, indicating the formation of cross-linking (see **Figure 5a**). An increase in cross-linking density correlates with a rise in the storage modulus. Rheology measurements demonstrated that the solution transitioned to a gel state within thirty seconds after initiating the measurement, and the storage modulus stabilized after five minutes, suggesting minimal further changes in cross-linking density. In the first five minutes, the change was significant, so we monitored thermal conductivity at 10-second intervals during this phase. After five minutes, the sample interval is changed to 20 seconds. Compared with the gelatin solution without initializing the gelation process, a noticeable increase in thermal conductivity was detected in the first four minutes following the mixing of the solution (**Figure 5b**). After five minutes, the thermal conductivity stabilized. In part 10 of SI, we demonstrate more measurement data of the gelation process of the hydrogel, and they all show similar increasing trends.

A comprehensive theoretical model is established to describe the phenomena that take place during the gelation process. In the model, we consider the heat transfer due to heat conduction, which governs the temperature distribution in and across different materials. Meanwhile, crosslinking dynamics are also important as they determine the temporal evolution of crosslinking density which has a great impact on the local material properties. The detail of the model is described in the SI. By combining the two parts, we simulate the thermal conductivity change in the hydrogel gelation process, and the result is consistent with the experimental findings (as shown in **Figure 5b**).

Before the experiment, we carefully tested the properties of the sensor, such as temperature sensitivity and repeatability, and we chose the sensor with the best quality. We also calibrated the thermal conductivity measurement and provided proof of the concept measurement. These guarantee the accuracy of our measurement. During the gelation process, cross-linking forms, which causes an increase in thermal conductivity. [11] The increasing trends of the measured result demonstrate the overall increase in the cross-linking density. For most hydrogels, the broad distribution of the molecular weight of the polymers and their nonuniform mixing before gelation will cause dramatic variation in local cross-linking density. [10] Also, the hydrogel in the gelation process is in a 'dynamic' status. The polymers are continually cross-linking, which causes the cross-linking density around the sensor to change constantly. Different from the traditional thermal conductivity measurement method, which records the 'average' result over a relatively large area, our measurement is for localized thermal conductivity. The spatial and temporal uniformity of the hydrogel is recorded as the variation of the measurement. Considering the sensor's size is comparable to the dimensions of the hydrogel network (detailed in the SI), it's conceivable that during cross-linking formation, changes in the distance between the sensor and the network structure could affect microscale thermal conductivity as it will be influenced by the effect of the interface. [19] Limited by the method's principle, we cannot simultaneously measure multiple points to obtain a statistical distribution of results, which we think is the biggest shortcoming of our method.

Microscale and nanoscale thermal conductivity of the hydrogel is still a brand-new field. Only one study has demonstrated the microscale thermal conductivity based on pure simulation. Although they show a similar trend that thermal conductivity will increase with cross-linking concentration, which is the same as the macroscale study, [11] the value they get is higher than the macroscopic one. This indicates the demand for more research on the difference between bulk scale and micro scale. The new method we developed provides a new tool for studying localized thermal conductivity and can benefit this area.

## 3. Conclusion

In summary, we introduced an innovative thermal conductivity measurement scheme supported by a newly developed ND@silicon nanopillar sensor, which is easy for mass production. This method uses two lasers with distinct focal points on the sensor, offering a significant advantage over traditional methods by minimizing crosstalk between the heating and sensing lasers and providing greater flexibility in selecting laser wavelengths. For the first time, we have demonstrated changes in thermal conductivity throughout the gelation process using our cutting-edge sensor and methodology. Our research offers a novel approach for localized micro-scale thermal conductivity measurements, serving as a potent tool for investigating material thermal properties. Additionally, the concept of decoupled sensing may inspire advancements in quantum sensing by reducing crosstalk in experiments involving multiple lasers. This work not only enhances our understanding of material thermal properties but also sets the stage for future developments in quantum sensing and related fields.

**4. Experimental Section/Methods**

*Morphology Characterization*: The surface morphology of the sensor was measured by NanoWizard 4XP BioScience atomic force microscope (Bruker, USA) in the air under ambient conditions. The measurements were performed under Quantitative Imagine (QI) Mode using commercially available ScanAsyst-Air AFM probes (Bruker, USA) with 0.4 $Nm^{-1}$ spring constant.

*Optical Characterization*: The confocal images were measured by a lab-built confocal microscope. The sample scan confocal microscopy was based on a customized piezo scanning stage (Core Tomorrow, Harbin). In the excitation path, a continuous-wave 532 nm laser (MGL-III-532, CNI) was used to excite the SiV Centers through an air objective (40X 0.95NA, UPLSAPO40X2, Olympus). In the heating laser path, a continuous-wave 637 nm laser (MGL-III-637, CNI) was used, and a 650 nm cut-on long pass dichroic mirror (DMLP650, Thorlabs) was used to import the laser into the light path. A single photon counting module (SPCM-AQRH-44-FC, Excelitas Technologies) was used to collect the fluorescence signal, and a data acquisition card (PCIe-6321, National Instruments) was used to read the signal of the SPCM. A 650 nm long pass filter (FELH0650, Thorlabs) was used to eliminate the 532 nm scattering light.

The spectrum measurement was conducted on the same confocal microscope. A 90:10 split Fiber Optic Coupler (TW805R2F1, Thorlabs) was used to import 90 percent of the light into the spectrometer (IsoPlane-81, Teledyne Princeton Instruments) for spectrum measurement. The integration time was 1 second, except for the shot noise measurement.

*Hydrogel Synthesis*: The hydrogel was made of gelatin solution and glutaraldehyde solution. The gelatin was diluted by phosphate-buffered saline (PBS) buffer (PH 7.4), and the concentration of the gelatin was fixed at 4% (w/w), the concentration of glutaraldehyde was 50% in $H_2O$. The gelatin solution and glutaraldehyde solution were mixed with a volume ratio of 10:1.

*Rheology Measurement*: The mechanical property, storage modulus (G'), and loss modulus (G'') of the hydrogel gelation process and gelatin solution were measured by a rheometer (MCR 302, Anton Paar) using a parallel plate with a 25mm diameter (PP25) in single frequency model. In the hydrogel gelation process measurement, 500μL 4% gelatin solution and 50μL 50% glutaraldehyde solution were mixed and quickly loaded on the platform. In the gelatin solution measurement, 500μL 4% gelatin solution was loaded on the platform. The tests were performed immediately after the samples were loaded. Measurements were conducted at room temperature with a frequency of 0.5 Hz and a constant 0.1% strain. The sample rate was 12 samples per minute.

*Hydrogel porous structures measurement*: The hydrogel was frozen in liquid nitrogen and then dried in a vacuum freeze dryer. The scanning electron microscope figures of the hydrogel were captured by Hitachi S-3400N.

**Acknowledgments**

Z.C. acknowledges the financial support from the National Natural Science Foundation of China (NSFC) and the Research Grants Council (RGC) of Hong Kong Joint Research Scheme (Project No. N_HKU750/23) and the Health@InnoHK program of the Innovation and Technology Commission of the Hong Kong SAR Government.

Y. L. thanks the financial support from the Research Grants Council (Project No. GRF/17210520), the Health@InnoHK program of the Innovation and Technology Commission of the Hong Kong SAR Government, and the National Natural Science Foundation of China (Project No. 12272332).

The authors would like to thank Dr. Yong Hou of HKU for his help with the preparation and revision of this paper.

**Conflict of Interest**

The authors declare no conflict of interest.

# Supporting Information

**A diamond heater-thermometer microsensor for measuring localized thermal conductivity: a case study in gelatin hydrogel**

*Linjie Ma[#], Jiahua Zhang[#], Zheng Hao, Jixiang Jing, Tongtong Zhang, Yuan Lin, Zhiqin Chu\**

## 1. Tuning Shape of the Sensor

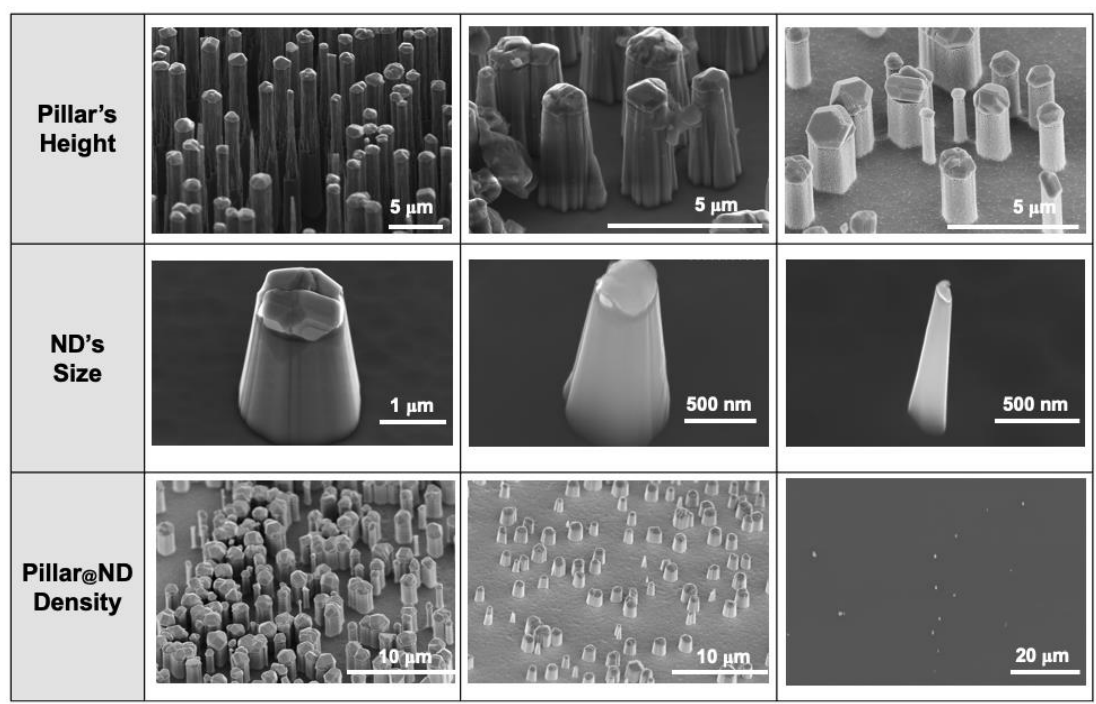

**Figure S1.** The sensors with different geometric shapes.

In the Pillar's Height line, the Inductively Coupled Plasma (ICP) etching time is changed. From left to right, the ICP etching time is 800s * 4, 800s, 300s.

In the ND's Size line, the Chemical vapor deposition (CVD) processing time is changing, From left to right, the CVD processing time is 80mins, 35mins, 0mins.

In the Pillar@ND Density line, the spin coating parameter is changed. From left to right, the condition is:

Left: 50nm particle size diamond powder water dispersion (1mg/mL), plus DMSO + acetone + anhydrous ethanol, equal volume mixed solution, ultrasonicate for 2 hours before use. Spin coating parameters: low speed 500 r/min, add 3 drops of the dispersion within 15 seconds, high speed 4500 r/min spin coating and drying for 110 seconds, repeat 8 times.

Middle: 50nm particle size diamond powder water dispersion (1mg/mL), ultrasonicate for 2 hours before use. Spin coating parameters: low speed 500 r/min, add 3 drops of the dispersion within 15 seconds, high speed 4500 r/min spin coating and drying for 110 seconds, repeat 8 times.

Right: 50nm particle size diamond powder water dispersion (1mg/mL), plus DMSO + acetone + anhydrous ethanol, equal volume mixed solution, ultrasonicate for 2 hours before use. Spin coating parameters: low speed 500 r/min, add 1 drop of the dispersion within 15 seconds, high speed 5000 r/min spin coating and drying for 110 seconds.

## 2. Characterization of the Sensor

### a. Typical Morphology of the Sensor

a) 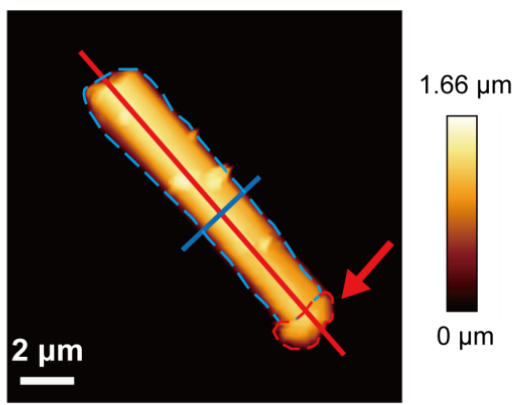

b) 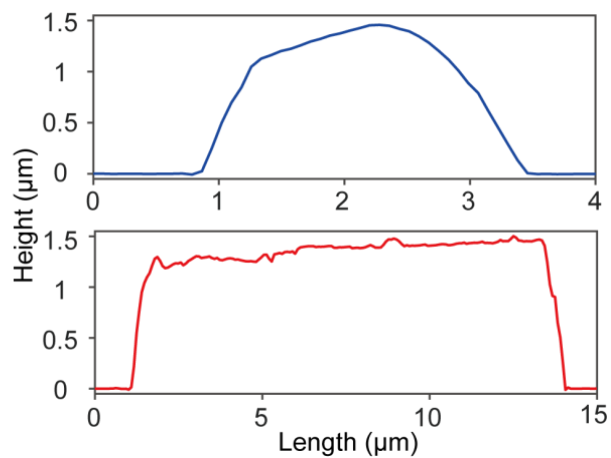

**Figure S2.** a. The AFM measurement result of the sensor. The red arrow indicates the location of the diamond particle. The region inside the red dashed line is the diamond particle, and the region inside the blue dashed line is the silicon pillar. The red and blue solid lines are the sample region of the line profile. b. The line profile of the AFM measurement result. The blue and red lines correspond to the blue and red lines in Figure a.

### b. Statistic of the sensor

Diameter:

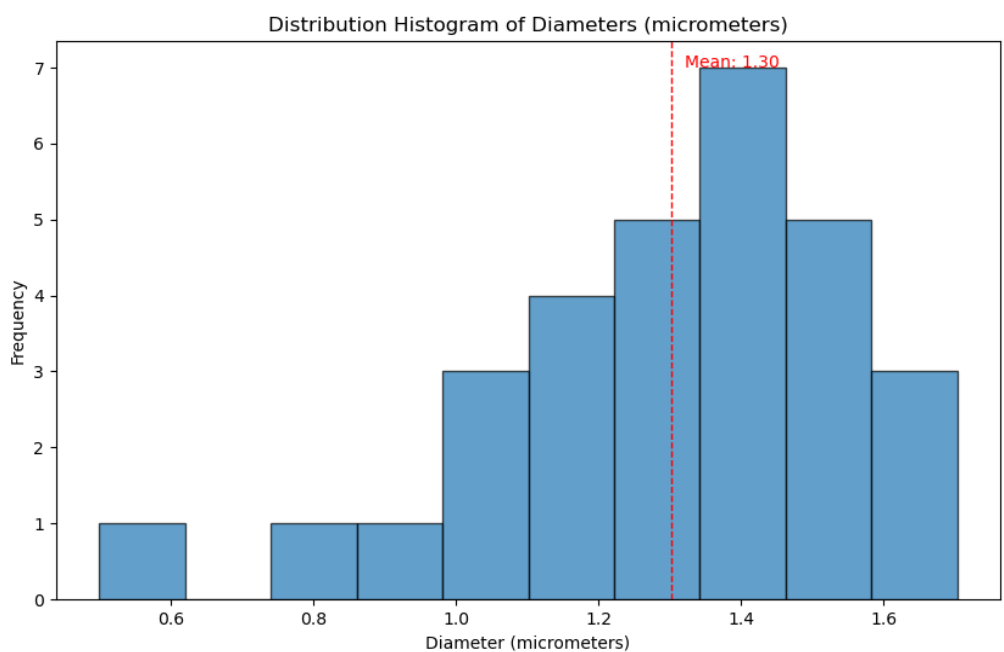

**Figure S3.** Statistic result of the diameter of the sensors

Mean: 1.30 μm

Standard Deviation: 0.27 μm

The result is calculated based on 30 samples.

Length:

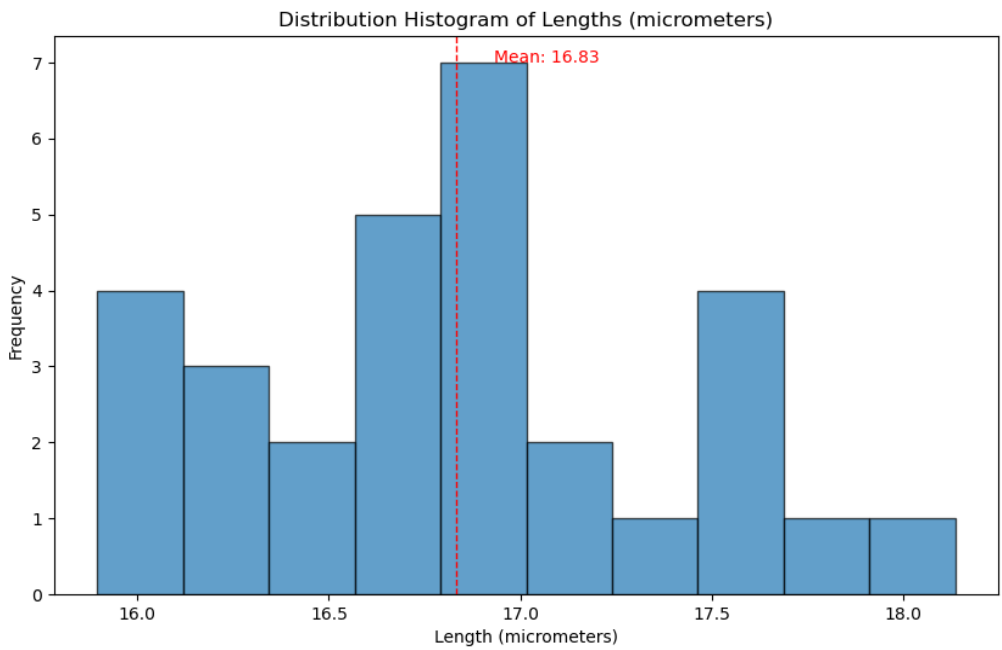

**Figure S4.** Statistic result of the length of the sensors

Mean: 16.83 μm

Standard Deviation: 0.56 μm

The result is calculated based on 30 samples.

## 3. Optical Property of the Sensor
### a. Typical spectrum of the sensor

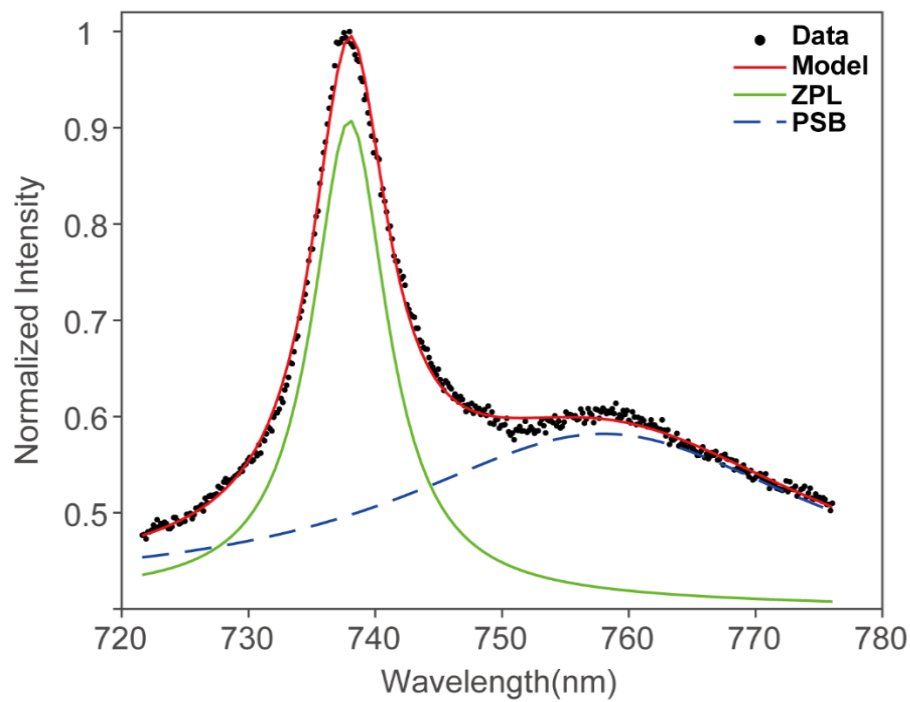

**Figure S5.** The spectrum of the diamond particle's fluorescence and the fitting model of the data, including separated Zero Phonon Line (ZPL) and phonon sideband (PSB). The ZPL located around 738 nm and the PSB located around 760 nm indicate the existence of SiV centers.

b. Statistic of the sensitivity

| Number | Sensitivity ($K \cdot Hz^{-\frac{1}{2}}$) |
|---|---|
| 1 | 1.10 |
| 2 | 2.71 |
| 3 | 3.50 |
| 4 | 2.85 |
| 5 | 3.30 |
| 6 | 1.34 |
| 7 | 3.03 |
| 8 | 1.60 |
| 9 | 1.40 |
| 10 | 2.60 |

Mean: $2.34 \text{ K} \cdot \text{Hz}^{-\frac{1}{2}}$

Standard Deviation: $0.85 \text{ K} \cdot \text{Hz}^{-\frac{1}{2}}$

## 4. Simulation model of laser heat effect on sensor

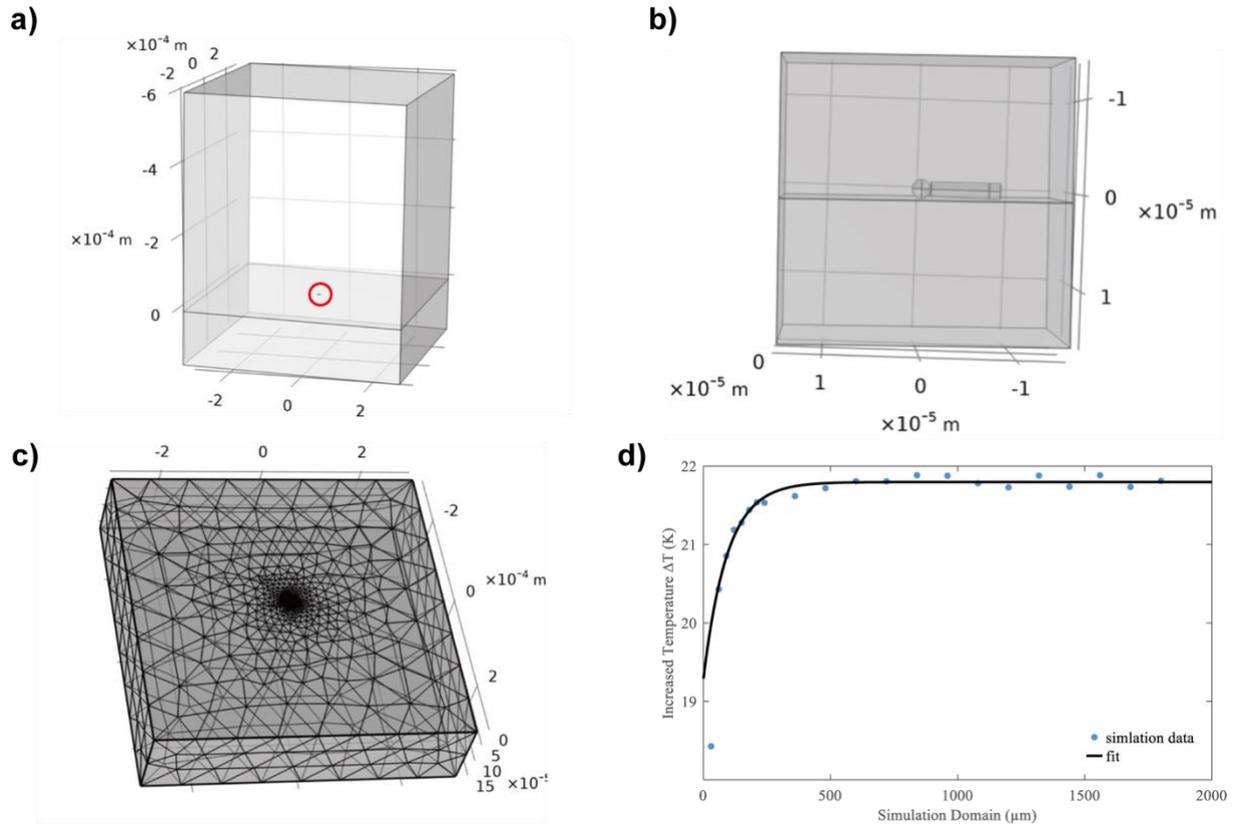

**Figure S6. a.** The geometric model of the simulation. The red circle indicates the location of the sensor. **b.** Zoom in view of the sensor. **c.** The mesh grid of the simulation model. **d.** The relationship between the simulation domain and the simulated increasing of temperature at the center of the nanodiamond

We build a simulation model to investigate the thermal effect of the heating laser on the sensor. As illustrated in **Figure S6. a.** and **b.**, the model consists of two parts: a 2 µm diameter sphere representing the diamond particle, and a Frustum of a cone representing the silicon pillar, with a diameter of 1.5 µm at the upper surface and 1.7 µm at the lower surface. These sizes are chosen based on the actual dimensions of the sensor. The substrate (cover glass) is located below the diamond, with a simulation area of 600 µm (L) * 600 µm (W) * 150 µm (H). The area covering the diamond is filled with different media (air, water, ethanol) based on the experiment conditions, and their corresponding simulation area is a 600 µm (L) * 600 µm (W) * 600 µm (H) cube. The nanodiamond is placed at the center of the cube. **Figure S6. c**. shows the mesh grid. To ensure size-invariant result, we choose simulation dimensions of 600 µm, as shown in **Figure S6.d**. For the boundary condition, as the experiment is performed under room temperature conditions, the temperature at the boundary has been fixed to the room temperature in the simulation.

In the simulation, the bottom of the silicon pillar, which is illuminated by the red laser, is considered as the heat source $Q_0 = P_0/V$, where $P_0$ is the heat power and $V$ is the heat source volume. The volume of the heat source is determined by the illuminated area of the Gaussian laser. The beam waist radius $\omega(z)$ at different transfer distance $z$ is related to the beam waist radius $\omega_0$ as follows:

$$\omega(z) = \omega_0 \sqrt{1 + \left(\frac{z\lambda}{\pi\omega_0^2}\right)^2}$$

As our laser is shining at the middle height of the pillar, we choose the thickness to be twice the laser beam waist radius at half height $\omega(z = 0.85\ \mu m)$. Furthe, we simulate the condition of Gaussian laser transmission through the pillar, as shown in the **Figure S7.b**. The laser can transmit through the entire bottom of the pillar, so we choose the height of the heat source to be the diameter of the pillar bottom, which is 1.7 µm.

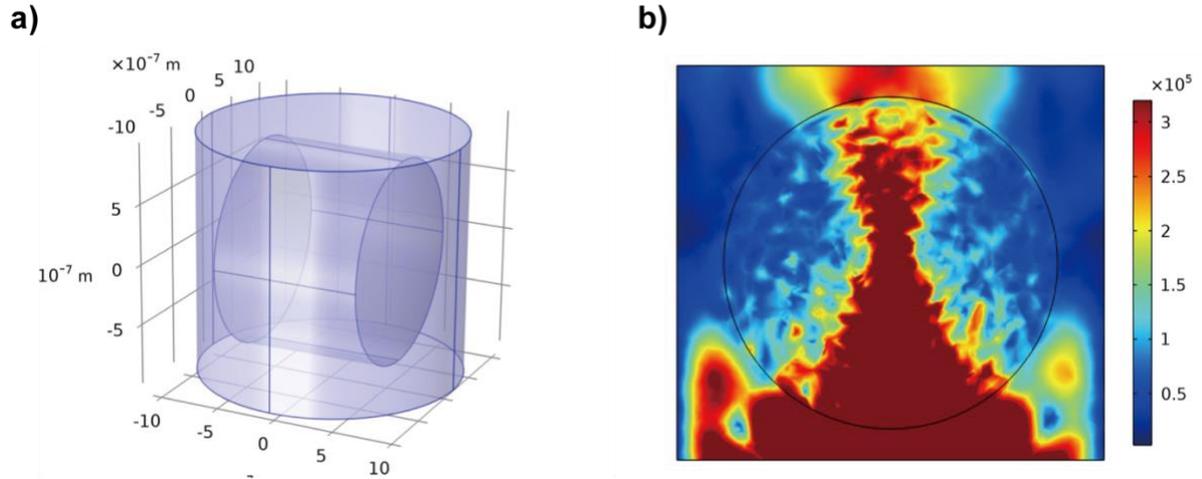

**Figure S7. a.** The model of the simulation. **b.** The electrical field simulation for Gaussian laser transmission through the pillar. The middle circle is the silicon pillar, and the outside is air.

To determine the heat energy generated from photothermal effect, we connect the heat power $P_0$ with laser intensity $I$ through $P_0 = \sigma_{abs} * I$. The light intensity can be derived from the formula:

$$P = \iint I e^{-2(x^2+y^2)/\omega_0^2}\, dxdy = I\frac{\pi\omega_0^2}{2} \Rightarrow I = \frac{2P}{\pi\omega_0^2}$$

Where $P$ is the laser power. The waist radius $\omega_0$ is 350nm in air, the waist radius in other materials can be calculated from the corresponding refractive index. The absorption cross section $\sigma_{abs}$ can be calculated by the electrical field simulation when a plane Gaussian laser transmit through the above-mentioned heat source volume. The simulation result for the absorption cross section is around $\sigma_{abs} = 1.72 \cdot 10^{-14} \ m^2$. We use this parameter to simulation the temperature rising after laser heat on the bottom of silicon pillar.

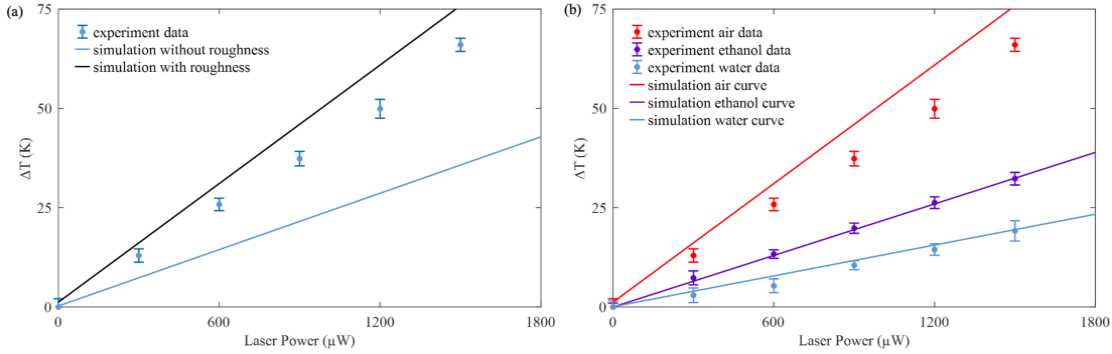

**Figure S8. a.** The surface roughness influence for laser heat effect. **b.** The simulation and experiment results for laser heat effect under three conditions (air, water, ethanol).

The reference study by Setoura et al. [1] highlights the significance of accounting for substrate surface roughness and avoiding the use of point contact method for particle-surface interaction in simulations. These factors can lead to inaccurate temperature predictions compared to experimental data. We present our simulation results with and without surface roughness for the air condition in **Figure S8**, which aligns with the findings of the reference. As a result, we consider the influence of the average surface roughness of the substrate, which is set to be 5nm, in our simulations.

In the simulation results for air/glass, we observed higher temperatures compared to the experimental data. We attribute this difference to the adsorbed water layer effect, as reported by Saito et al. [2] In ambient atmosphere, the substrate surface forms an adsorbed water layer with a thickness of several nanometers, which can influence the temperature readings. To account for this effect in our simulation, we assigned a lower surface roughness value of 3nm to air, which enhances the thermal energy transfer.

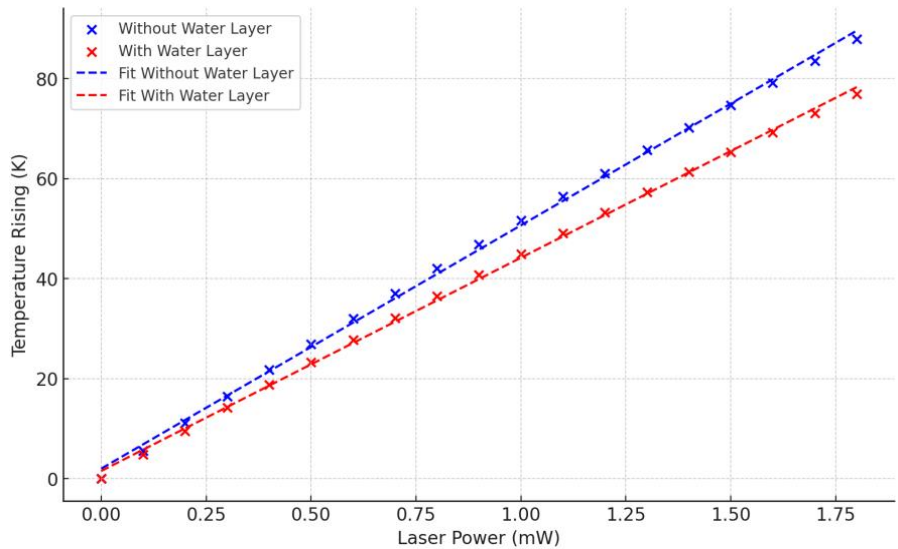

**Figure S9.** The simulation result and the liner fitting of the temperature rising in air with and without considering the water layer.

The simulation aims to show that within a certain range, the increase in temperature is linearly related to the power of the heating laser. From **Figure S9**, we can see that the simulation result shows that although taking the water layer into consideration does alter the specific values of the temperature increase, the linear relationship remains consistent. Therefore, based on our experimental results, we include the effect of the water layer in the simulation for correction. The detailed impact of the water layer is complicated. However, the existence of the water layer at the interface has already been widely accepted in other fields. For example, in AFM experiments, the adhesion force between the tip and the substrate in the air is generally greater than in liquid, which is due to the capillary force caused by the water layer on the surface in air. [3]

Simulation parameter:
Table of material thermal conductivity:

| Material | Thermal Conductivity ($WK^{-1}m^{-1}$) |
| --- | --- |
| SiO2 Glass | 1.38 |
| Diamond | 1200 |
| Silicon | 149 |
| Air | 0.026 |
| Water | 0.6 |

| Ethanol | 0.16 |

In the simulation, the mesh grid is minimum 1.6 nm.

The simulation takes 3 mins with 2.35 GB memory need.

Computer (16 GB memory, i5-1135G7@2.4 GHz).

## 5. Derivation of Thermal Conductivity from Temperature Measurement

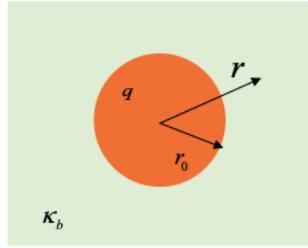

**Figure S10.** The schematic of the model.

A sphere object serves as the heat source, with the heat generated by the absorption of laser power. The steady increase in temperature under continue wave (cw) laser illumination in detector/heat (1D model) can be calculated using the equation $\Delta T = \frac{\sigma_{abs} I}{4\pi R \kappa_b}$ [4]. The laser intensity $I$ can be obtained from the laser power $P$, using the method mentioned earlier in the simulation section. We can observe an inverse relationship between the laser heat slope $k_{LH}$ and the environmental thermal conductivity.

$$k_{LH} = \frac{d\Delta T}{dP} \propto \frac{Const}{\kappa_b}$$

When extended to 3D models in the real application, we expect to observe an increase in temperature with power will increase when thermal conductivity is lower. To explore the actual relationship between environmental thermal conductivity and laser heat slope in the 3D pillar model (our experiment setup), we use the above mentioned simulation model in COMSOL and obtained simulation results, which are presented in **Figure S11.** In our initial thought, we did not factor in surface roughness during the simulation process. Therefore, we conducted the simulation under normal point contact conditions for the interaction between the pillar and surface. The simulation results were fitted using a power law relationship with a negative power well, resulting in an R-squared value of 0.9994. The fitting curve for the laser heat slope is expressed as $k_{LH} = A * \kappa^B + C$, where A=0.001066, B= -0.8615, and C=0.001457.

Although the actual conditions may be more complex than what was simulated, we still believe that there is a general relationship between the laser heat slope and thermal conductivity, which can be expressed as follows:

$$k_{LH} = \frac{dT}{dP} = A * \kappa^B + C$$

Where B < 0. A, B, C are the parameters to be determined by the experiment.

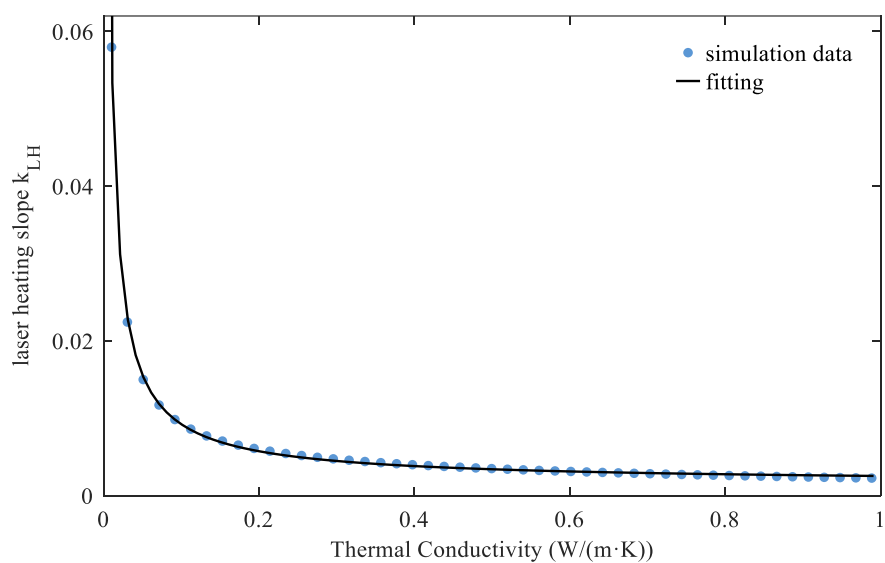

**Figure S11.** The simulation model for the relationship between laser heat slope and environment thermal conductivity

## 6. Validation of thermal conductivity measurement

To validate the thermal conductivity measurement method we proposed, we measured the thermal conductivity of the ethanol solutions with different concentrations. We compared the data obtained from the experiment with the theoretical calculation data. The following figure and table show both the experiment and theoretical data.

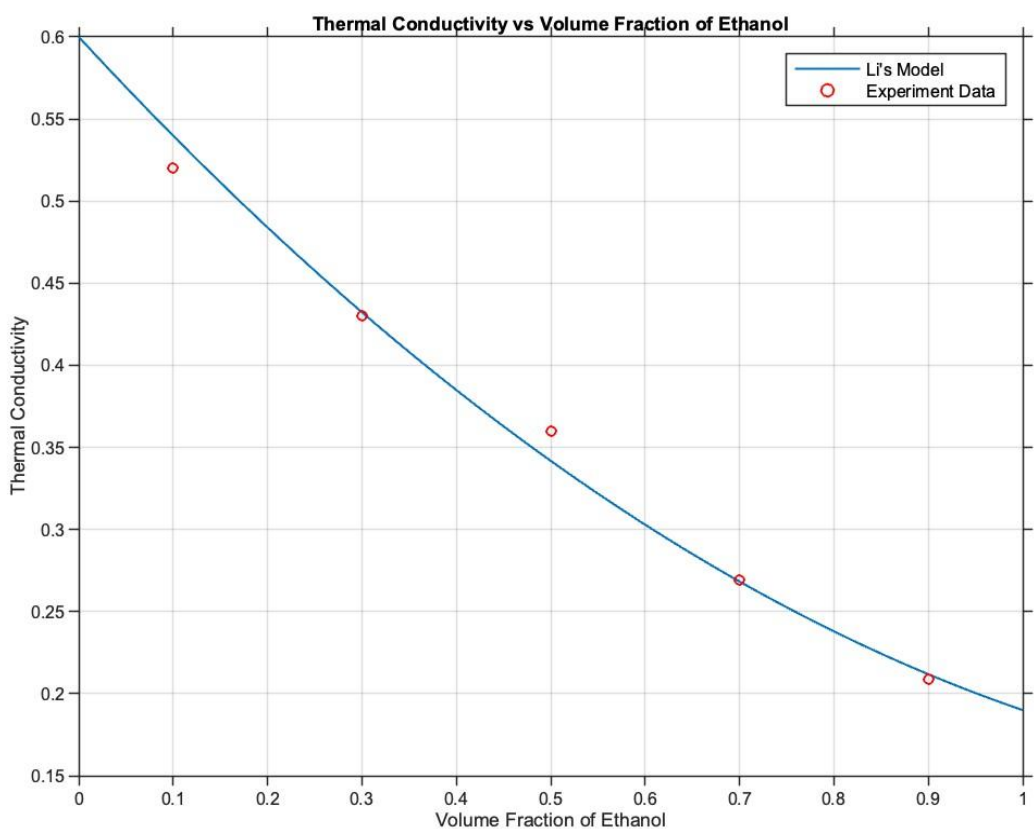

**Figure S12.** The theoretical calculation result and experiment measurement result of the Ethanol solution with different volume fractions of the ethanol.

| Material | 10% Ethanol | 30% Ethanol | 50% Ethanol | 70% Ethanol | 90% Ethanol |
|---|---|---|---|---|---|
| Measured | 0.52 | 0.43 | 0.36 | 0.27 | 0.21 |
| Theory | 0.540 | 0.433 | 0.342 | 0.268 | 0.212 |

**Table S1.** The theoretical calculation result and experiment measurement result of the Ethanol solution.

The simulation is based on the method of Li.[5] The mixture rule for the thermal conductivity of liquids is:

$$\lambda_{mix} = \sum_{i=1}^{n} \sum_{j=1}^{n} \frac{2\Phi_i \Phi_j}{\lambda_i^{-1} + \lambda_j^{-1}}$$

with

$$\Phi_i = \frac{\tilde{x}_i v_{liq,i}}{\sum_{j=1}^{n} \tilde{x}_j v_{liq,j}}$$

Where $\tilde{x}_i$ is the mole fraction of the component, $v_{liq,i}$ is the mole volume of the component, $\lambda_i$ is the thermal conductivity of the component.

## 7. Measurement Accuracy Analysis

Here we analyze the accuracy of the thermal conductivity measurement.

The thermal conductivity is calculated from the slope of localized temperature rising and heating laser power, k. As derives in part 5, the relationship is:

$$k = A * \kappa^B + C$$

So the thermal conductivity $\kappa$ can be written as:

$$\kappa = \sqrt[B]{\frac{k - C}{A}}$$

Based on the Propagation of uncertainty:

$$\sigma_\kappa = \sqrt{(\frac{\partial f}{\partial k})^2 \sigma_k^2}$$

$$= \sqrt{(\frac{1}{AB}\left(\frac{k-C}{A}\right)^{\frac{1}{B}-1})^2 \sigma_k^2}$$

Similarly, the formula for the calculation of k is:

$$k = \frac{y_2 - y_1}{x_2 - x_1}$$

So, the uncertainty of k is:

$$\sigma_k^2 = \frac{\sigma_{y_1}^2 + \sigma_{y_2}^2}{(x_2 - x_1)^2}$$

Based on this formula, the calculated standard deviation of first time three solutions with different ethanol fraction are:

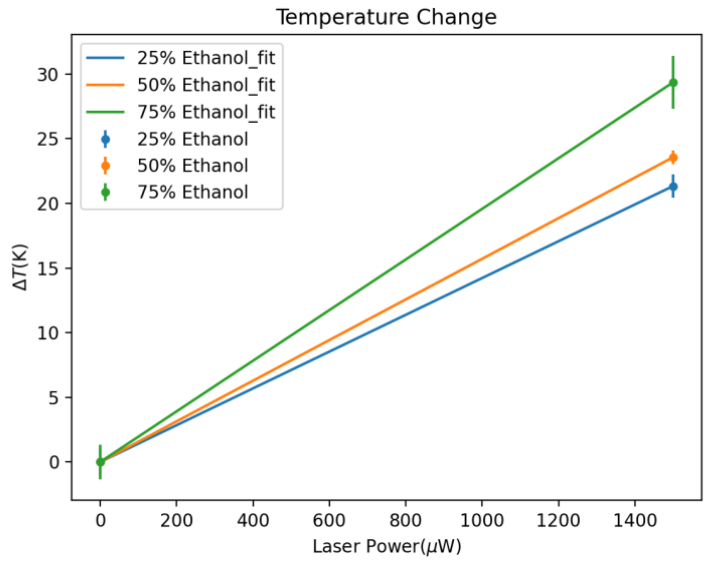

$$\sigma_{25\% \, Ethanol} = 0.061 W m^{-1} K^{-1}$$
$$\sigma_{50\% \, Ethanol} = 0.048 W m^{-1} K^{-1}$$
$$\sigma_{75\% \, Ethanol} = 0.048 W m^{-1} K^{-1}$$

The calculated standard deviation of second time five solutions with different ethanol fraction (the data demonstrated in the main text) are:

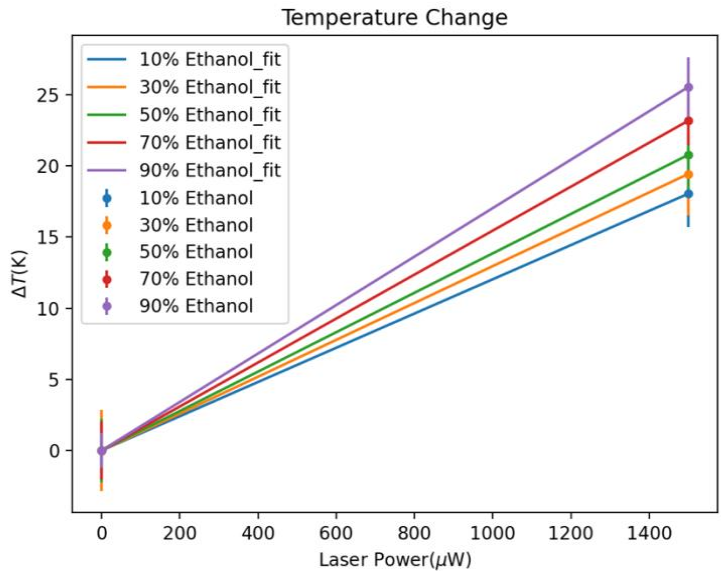

$$\sigma_{10\% \, Ethanol} = 0.252 W m^{-1} K^{-1}$$
$$\sigma_{30\% \, Ethanol} = 0.248 W m^{-1} K^{-1}$$
$$\sigma_{50\% \, Ethanol} = 0.160 W m^{-1} K^{-1}$$

$$\sigma_{70\% \, Ethanol} = 0.081 Wm^{-1}K^{-1}$$
$$\sigma_{90\% \, Ethanol} = 0.051 Wm^{-1}K^{-1}$$

## 8. Simulation of the Hydrogel Gelation Process

To build a model that can couple heat transfer due to heat conduction and crosslinking dynamics, 2 links need to be established first. The first link is the crosslinking dynamics expressed with the crosslinking density. Equipping our model with it will enable the crosslinking density to be solved explicitly given the temperature. The second one is the mechanism on how the crosslinking density affects the thermal conductivity of the gelatin. This link is vital in that the thermal conductivity it provides is an indispensable parameter in heat transfer modelling. For the following section, the above 2 links will be discussed respectively, then the framework of our theoretical model will be introduced.

### 8.1. Crosslinking dynamics

Crosslinking is the chemical reaction between the polymer and the crosslinker, where the crosslinker simultaneously bonds with different parts of the polymer to change the polymer's inner structure. In this study, the polymer is gelatin and the crosslinker is glutaraldehyde. The simplified equation of this crosslinking reaction is given as follows,

$$2C_0 + G \Longrightarrow 2C_1 \quad (1)$$

where $C_0$ and $C_1$ refer to the crosslinkable site and crosslinked site within the gelatin, and G refers to glutaraldehyde. Typically, the reaction rate of equation (1). can be assumed to be dependent on the concentrations of its reactants, and a general form of this reaction rate is given as follows,

$$\frac{dc_1}{dt} = k_{cl} c_0^m c_g^n \quad (2)$$

where $c_0$ and $c_1$ are the concentrations of the crosslinkable sites and crosslinked sites within the gelatin, $c_g$ is the concentration of the glutaraldehyde, m & n are the orders of dependence, and $k_{cl}$ is the temperature-dependent rate constant. Following the treatment from Robinson's work, [6] the first-order reaction assumption is adopted here and equation (2). has been transformed into the following form,

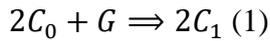

$$\frac{dc_1}{dt} = k_{cl} c_0 c_{g0}^n \quad (3)$$

where m is set to 1 and $c_g$ is replaced with $c_{g0}$, the initial concentration of the glutaraldehyde. The logic behind the replacement of $c_g$ can be roughly explained by both the fast refill of the crosslinker due to its ability to move freely across the polymer, and the relative abundance of the crosslinker in our study. To incorporate the crosslinking density ($\rho_{cl}$) into the model, the relation $\rho_{cl}=c_1/(c_0+c_1)$ is utilized to further transform equation (3). into the following form,

$$\frac{d\rho_{cl}}{dt} = k_{cl}(1 - \rho_{cl})c_{g0}^n \quad (4)$$

If rearranged to the integral form, taken the integration then taken the logarithm on both sides of the equation, equation (4). will be turned into the following equation,

$$\ln t_g = \ln \frac{-\ln(1-\rho_g)}{k_{cl}} - n \ln c_{g0} \quad (5)$$

where $t_g$ is the gelation time, $\rho_g$ refers to the crosslinking density at the time of gelation. What equation (5). implies is that, under the assumptions of our model, if other conditions of the crosslinking reactions remain fixed, $\ln(t_g)$ declines linearly with $\ln(c_{g0})$. Moreover, fed by the experimental data in Robinson's work [1], linear regression method is adopted to obtain the values of the terms $-\ln(1-\rho_g)/k_{cl}$ and n in equation (5). (0.25 and 2 respectively). While n can be utilized by direct substitution into eq. 4, $k_{cl}$ is still undetermined. The Arrhenius equation gives,

$$k_{cl} = k_{cl0} e^{-\Delta E/RT} \quad (6)$$

where $k_{cl0}$ is the temperature-independent constant factor, $\Delta E$ is the activation energy which, according to Hopwood et al, [7] is 72 kJ·mol$^{-1}$ on average for glutaraldehyde-related crosslinking reactions, R is the universal gas constant, and T is the absolute temperature. So, with $-\ln(1-\rho_g)/k_{cl}$ determined, for any specified $\rho_g$, a corresponding $k_{cl0}$ can be calculated, then enable the calculation of $k_{cl}$. However, due to the limitation on the measurement of the crosslinking density, the exact value of $\rho_g$ has not been mentioned in any literature within our scope. To make up for this conundrum, $\rho_g$ is chosen to act as a fitting parameter in our theoretical model, whose value will be updated by comparing the theoretical result against the experimental one.

To sum up, with the combined utilization of equation (4). and equation (6)., the crosslinking dynamics can be described given any specified temperature distribution.

### 8.2. Thermal conductivity calculation based on crosslinking density

Using the network model presented by Eiermann et al, Yamamoto deduced the theoretical relationship between the crosslinked polymer's thermal conductivity and its crosslinking density. [8] According to Yamamoto, this relationship can be roughly estimated using the following expression,

$$\frac{k_c}{k_0} = 1 + \frac{P^2 - P}{2P+1} \rho_{cl} \quad (7)$$

where $k_0$ and $k_c$ are the intrinsic thermal conductivity and crosslinking-dependent thermal conductivity of the polymer, $P = k_e/k_w$ is the ratio of the thermal conductivity through the covalent bond to that through the Van der Waals bond. In Yamamoto and Kambe's follow-up work, [9] the calculated results based on equation (7). agreed fairly well with the measured results, thus equation (7). is adopted here to calculate the thermal conductivity of the gelatin under the influence of crosslinking. According to Yamamoto [5], P can be further approximated to be 10, so the version of equation (7). used in our theoretical model can be simplified as follow,

$$\frac{k_{gel}}{k_{gel0}} = 1 + 4.3\rho_{cl} \quad (8)$$

where $k_{gel0}$ is the intrinsic thermal conductivity of the gelatin which, according to Sakiyama et al, [10] is measured to be around 0.28 $WK^{-1}m^{-1}$ across a wide range of temperature, and $k_{gel}$ is the true thermal conductivity of the crosslinked gelatin. To calculate the overall thermal conductivity of the water-gelatin mixture ($k_m$), the volume fractions of water ($\Phi_w$) and gelatin ($\Phi_{gel}$) also need to be specified, so that $k_m$ can be calculated using the following expression, [10]

$$k_m = \Phi_w k_w + \Phi_{gel} k_{gel} \quad (9)$$

where $k_w$ is the thermal conductivity of water, which is set to 0.59 $WK^{-1}m^{-1}$ in this study. Assuming that the gelatin mass and the water mass in the mixture are $m_{gel}$ and $m_w$ respectively, the volume fractions can be calculated as follows,

$$\Phi_{gel} = \frac{m_{gel}/\rho_{gel} + m_{gel}(R_s-1)/\rho_w}{V_m}, \Phi_w = 1 - \Phi_{gel} \quad (10)$$

where $\rho_{gel}$ and $\rho_w$ are the densities of the gelatin and water respectively, $R_s$ is the dimensionless swelling ratio of the gelatin, and $V_m$ is the total volume of the gelatin-water mixture. For simplicity, $R_s$ is considered to increase linearly with $\rho_{cl}$, and peak at 5 when $\rho_{cl}$ reaches 1 according to the measurement results of Anjali et al, [11] which can be expressed as,

$$R_s = 5\rho_{cl} \quad (11)$$

In conclusion, with equation 8-11 combined, the relationship between gelatin's thermal conductivity and its crosslinking density is established.

### 8.3. Modelling heat transfer coupled with crosslinking dynamics

Apart from establishing the above 2 links to connect the heat transfer process and crosslinking process, the individual modelling of these 2 processes was also carried out.

The heat transfer in our study is a heat-conduction-dominant process with negligible convection and radiation, so the framework of the solid heat transfer without radiation source is adopted to model the heat transfer process. The governing equation used in such a framework is presented as follows,

$$\rho C_p \frac{\partial T}{\partial t} + \nabla \cdot \mathbf{q} = Q, \mathbf{q} = -k\nabla T \quad (12)$$

where $\rho$, $C_p$, k and T are the density, specific heat capacity, thermal conductivity and absolute temperature of the medium, $\mathbf{q}$ is the heat flux by heat conduction, and Q is the heat source.

The crosslinking dynamics is considered by solving ODE defined with equation (4) and equation (6). As mentioned earlier, the crosslinking density at gelation ($\rho_g$) is used as a fitting parameter to determine the appropriate value for $k_0$, and here is how $\rho_g$ is determined. For any given $\rho_g$, a certain $k_{cl0}$ is calculated and utilized to calculate $k_{cl}$ with equation (6.) Then a full simulation of the heat transfer coupled with the crosslinking dynamics is carried out and outputs the temporal evolution of the crosslinking density. This temporal evolution will be qualitatively compared with the experimental results of the storage modulus in Figure 5. a. According to ideal rubber elasticity theory, the rubbery plateau tensile modulus, E ($\approx$ E′ when E″ is negligibly small), is proportional to crosslinking density. [12] Following this logic, if the patterns of $\rho_{cl}$ and E′ are identical to a certain degree, the current $\rho_g$ used to calculate $k_{cl0}$ can be considered appropriate, and the overall simulation process will stop and output the temporal evolution of the thermal conductivity. Otherwise, $\rho_g$ will be updated to start another full simulation.

The simulation based on the above modelling framework is realized using the COMSOL Multiphysics 6.0. The module "Heat Transfer in Solids" is employed to model the heat transfer process in the gelatin, and the module "Domain ODEs and DAEs" is employed to model the crosslinking process. The coupling between 2 processes is realized mainly by introducing 2 dependent variables, the temperature-dependent rate constant $k_{cl}$ and the crosslinking-dependent thermal conductivity $k_{gel}$.

## 9. Calibration of sensor and thermal conductivity measurement result of hydrogel gelation process

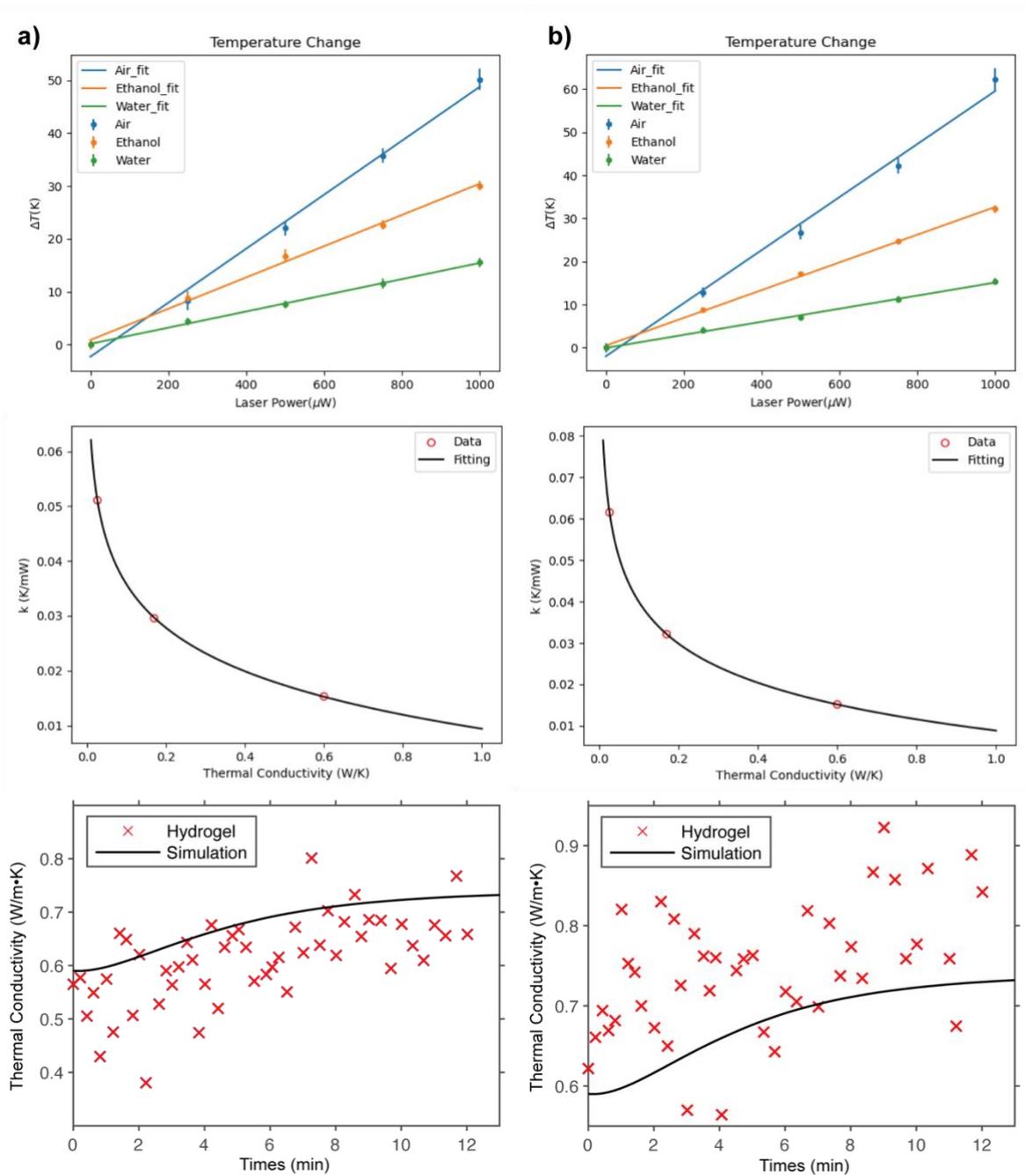

**Figure S13.** Sensor calibration and thermal conductivity measurement result of hydrogel gelation process.

10. **Hydrogel porous structures measurement**

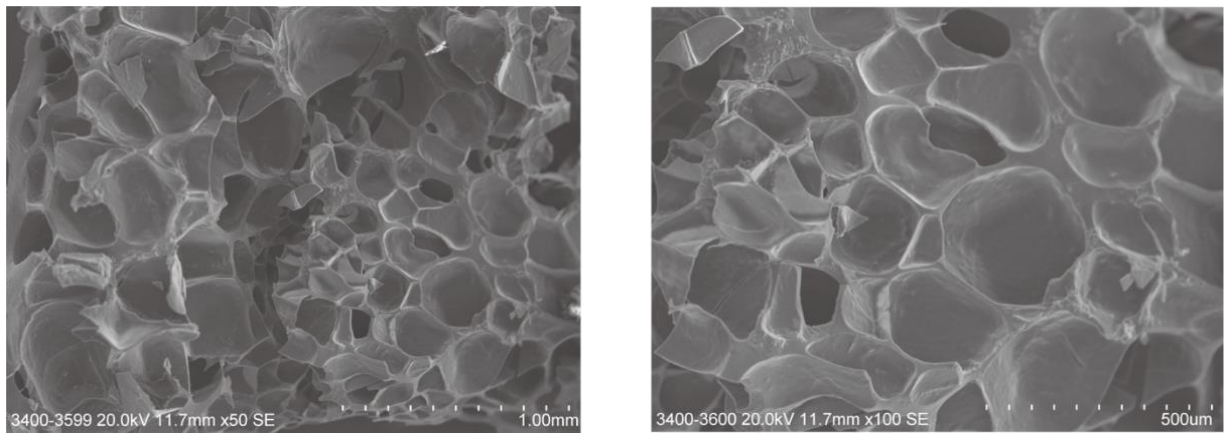

**Figure S14.** The SEM image of the freeze-dried hydrogel structure.